\documentstyle[11pt,fleqn,cite]{article}   
\catcode`\@=11
\@addtoreset{equation}{section}
\def\theequation{\arabic{section}.\arabic{equation}}
\def\appendix{\renewcommand{\thesection}{\Alph{section}}\setcounter{section}{0}
              \renewcommand{\theequation}
            {\mbox{\Alph{section}.\arabic{equation}}}\setcounter{equation}{0}}
\def\maketitle{\thispagestyle{empty}\setcounter{page}0\newpage
                \renewcommand{\thefootnote}{\arabic{footnote}}
                  \setcounter{footnote}0}
\renewcommand{\thanks}[1]{\renewcommand{\thefootnote}{\fnsymbol{footnote}}
               \footnote{#1}\renewcommand{\thefootnote}{\arabic{footnote}}}
\newcommand{\preprint}[1]{\hfill{\sl preprint - #1}\par\bigskip\par\rm}
\renewcommand{\title}[1]{\begin{center}\Large\bf #1\end{center}\rm\par\bigskip}
\renewcommand{\author}[1]{\begin{center}\Large #1\end{center}}
\newcommand{\address}[1]{\begin{center}\large #1\end{center}}
\def\dip{\smallskip Dipartimento di Fisica, 
                                Universit\`a di Trento, Italia}

\def\dinfn{\smallskip Dipartimento di Fisica, Universit\`a di Trento\\ 
                           and Istituto Nazionale di Fisica Nucleare,\\
                                   Gruppo Collegato di Trento, Italia}
\def\Idip{\address{\dip}}

\def\Idinfn{\address{\dinfn}}
\newcommand{\email}[1]{e-mail: \sl #1@science.unitn.it\rm}
\newcommand{\femail}[1]{\thanks{\email{#1}}}

\def\babs{\hrule\par\begin{description}\item{Abstract: }\it} 
\def\eabs{\par\end{description}\hrule\par\medskip\rm}
\renewcommand{\date}[1]{\par\bigskip\par\sl\hfill #1\par\medskip\par\rm}
\newcommand{\ack}[1]{\par\section*{Acknowledgments} #1} 
\newcommand{\s}[1]{\section{#1}}

\def\hs{\qquad}               
\def\nn{\nonumber}            
\def\beq{\begin{eqnarray}}    
\def\eeq{\end{eqnarray}}      
\def\at{\left(}               
\def\aq{\left[}               
\def\ct{\right)}              
\def\cq{\right]}              
\def\R{{\hbox{{\rm I}\kern-.2em\hbox{\rm R}}}}   
\def\H{{\hbox{{\rm I}\kern-.2em\hbox{\rm H}}}}   
\def\N{{\hbox{{\rm I}\kern-.2em\hbox{\rm N}}}}   
\def\C{{\ \hbox{{\rm I}\kern-.6em\hbox{\bf C}}}} 
\def\Z{{\hbox{{\rm Z}\kern-.4em\hbox{\rm Z}}}}   
\def\ii{\infty}                                  
\newcommand{\fr}[2]{\mbox{$\frac{#1}{#2}$}}      

\def\be{\beta}

\def\ep{\varepsilon}

\def\la{\lambda}

\def\si{\sigma}
\def\om{\omega}
\def\ph{\varphi}

\def\te{\vartheta}

\def\Ga{\Gamma}

\begin{document}

\def\Ls{L\:}
\preprint{UTF 402}

\title{Electromagnetic fields in Schwarzschild \\ 
and Reissner-Nordstr\"om
geometry. \\ Quantum corrections to the black hole entropy} 

\author{Guido Cognola\femail{cognola}}
\Idinfn
\author{Paola Lecca}
\Idip

\date{June 1997}

\babs
Using standard coordinates, the Maxwell equations in the 
Reissner-Nordstr\"om geometry are written in terms of a couple 
of scalar fields satisfying Klein-Gordon like equations. 
The density of states is derived in the semi-classical approximation
and the first quantum corrections to the black hole entropy is
computed by using the brick-wall model.  
\eabs

\s{Introduction}
In the last decade, many efforts have been made in order to understand 
the deep origin of the black hole entropy
\cite{york83-28-2929,thoo85-256-727,frol93-48-4545,suss94-50-2700}
first introduced by 
Bekenstein \cite{beke72-4-737,beke73-7-2333,beke74-9-3292}
(for a recent review see Ref.~\cite{beke94u-15}), but
widely embraced only after the Hawinkg's demonstration of black 
hole thermal radiation 
\cite{hawk74-248-30,hawk75-43-199,hart76-13-2188}. 
Some possible interpretations and methods of calculation have been proposed, 
but, at the moment, no one of them seems to be the true answer
\cite{thoo85-256-727,call90-340-366,sred93-71-666,dowk94-11-55,suss96-45-115,ghos97-78-1858}. 
Here we recall the 't~Hooft proposal \cite{thoo85-256-727,thoo96-11-4623}, 
in which the black hole entropy is identified with the entropy of the
quantum fields surrounding the black hole itself. Since the density 
of states approaching the horizon diverges, 
in order to avoid divergences in the entropy, 
he has to introduce a cutoff parameter of the order of the Planck's length, 
which is interpreted as the position of a ``brick wall''
(the brick wall model).
He also computed the contribution to the entropy of a Schwarzschild black hole
due to a scalar field using a semi-classical approximation. 
After this, quantum corrections to the Bekenstein-Hawking entropy, 
due to a scalar field, 
have been computed by different methods for Schwarzschild 
\cite{cogn95-12-1927,frol96-54-2711}
Reissner-Nordstr\"om 
\cite{deme95-52-2245,ghos95-357-295,cogn95-52-4548,ghos96-11-2933} 
and also for Kerr-Newman \cite{ho97u-216} black holes
(for a recent review on quantum corrections to black hole entropy see
for example Ref.~\cite{frol96-54-2711}).

Also in the case of scalar fields, due to technical difficulties, 
one has to make some suitable approximation.
Some authors directly consider the Rindler space, which can be considered 
as an approximation of the Schwarzschild case for very large mass.
In the Rindler space, the contribution to the entropy due to scalar 
and also higher spin fields have been considered in 
Refs.~\cite{kaba95-453-281,more96-54-7459,more97-55-3552}, 
where in particular it has been shown that, 
depending on the method of calculation used, 
the contribution of the electromagnetic field is not just twice 
the scalar one, but it contains some unexpected anomalous surface terms.

In the present paper we focus our attention on the electromagnetic
field in Reissner-Nordstr\"om back ground. 
We solve the Maxwell equations and then 
compute the contribution to the black hole entropy by using the brick
wall model and show that no anomalous term is present. 
All results are valid for the Schwarzschild black hole in the trivial
limit of vanishing charge.
Maxwell equations in Schwarzschild, Reissner-Nordstr\"om and also Kerr
metric are usually solved by using Newman-Penrose formalism 
(see for example Ref.~\cite{pric72-5-2419} and Ref.~\cite{chan83b}
for a complete treatment of solutions in such a formalism), 
which is not familiar to many readers. For this reason
here we prefer to use a more conventional method, which consists
in solving Maxwell equations for the electromagnetic potential
in standard coordinates in a suitable gauge.

The paper is organized as follows. In Section 2 we consider the
Maxwell equations for the electromagnetic potential
in the Reissner-Nordstr\"om back ground and show that
they reduce to the ones of a couple of
independent scalar fields, which we solve in the semi-classical 
approximation, following the 't~Hooft's original work 
\cite{thoo85-256-727,thoo96-11-4623}. 
In this way we easily derive the expected contribution to the 
Bekenstein-Hawking entropy in Section 3.

As usual we use natural units in which $G=\hbar=c=k=1$.

\s{Electromagnetic fields in Schwarzschild and 
Reissner-Nordstr\"om black holes}

Here we study the electromagnetic waves in the 
Reissner-Nordstr\"om background (the solutions in the Schwarzschild  
geometry will be obtained as a limiting case for $Q\to0$), 
that is the non-static solutions of the equations 
\beq 
\nabla_iF^{ij}=0\:,
\hs i,j=0,...,3\:,
\label{ME}
\eeq
in the metric $g_{ij}$ given by
\beq 
ds^2=-\at1-\fr{2M}{r}+\fr{Q^2}{r^2}\ct\:dt^2
+\at1-\fr{2M}{r}+\fr{Q^2}{r^2}\ct^{-1}\:dr^2+r^2\:d\si^2\:,
\eeq
$F^{ij}$ being the electromagnetic field strength, 
$\nabla_k$ the covariant derivative, $M$ and $Q$ the mass and the charge
of the black hole respectively and
finally $d\si^2$ is the metric on the unit sphere, 
which is usually written in polar coordinates $\{\te,\ph\}$, 
but for our purposes the (complex) stereo-graphic coordinates
are more convenient. Then we write
\beq 
d\si^2=d\te^2+\sin^2\te\:d\ph^2=\frac{4}{(1+z\bar z)^2}dz\:d\bar z\:,
\eeq
where 
\beq 
z=\frac{\sin\te e^{i\ph}}{1-\cos\te}\:,
\hs \bar z=\frac{\sin\te e^{-i\ph}}{1-\cos\te}\:.
\eeq
In such a coordinates the non-vanishing components of the metric read
\beq 
g_{00}\equiv g_{tt}&=&-\at1-\fr{2M}{r}+\fr{Q^2}{r^2}\ct\:,
\hs g_{11}\equiv g_{rr}=-\at1-\fr{2M}{r}+\fr{Q^2}{r^2}\ct^{-1}\:,\nn\\
g_{23}\equiv g_{z\bar z}&=&\frac{2r^2}{(1+z\bar z)^2}=g_{32}\equiv g_{\bar zz}\:.
\eeq

In terms of the electromagnetic potential $A_k$, Eqs.~(\ref{ME}) 
become
\beq 
\Box A_k-\nabla_j\nabla_k A^j=
\Box A_k-\nabla_k\nabla_j A^j-R_{kj}A^j=0
\eeq
and finally, after some calculations we can put them in the more useful form
\beq 
\Ls A_k=A_j\partial_k\Ga^j+2g^{rs}\Ga^j_{rk}\partial_sA_j
+\partial_k\nabla_jA^j\:,
\label{ECE}
\eeq 
where $\Ga^j=g^{rs}\Ga^j_{rs}$,
$\Box=g^{ij}\nabla_i\nabla_j$ is the Dalembertian's operator while
$\Ls$ represents the Dalembert's operator acting on functions,
that is
\beq
\Ls = \frac1{\sqrt{|g|}}\partial_i\sqrt{|g|}g^{ij}\partial_j\:.
\eeq

From Eq.~(\ref{ECE}), after straightforward calculations we get
\beq 
\Ls A_0&=&
-2\at\frac{M}{r^2}-\frac{Q^2}{r^3}\ct
(\partial_0A_1-\partial_1A_0)+\partial_0\nabla_jA^j
\:,\label{LA0}\\
\Ls A_1&=&-2\at\frac{M}{r^2}-\frac{Q^2}{r^3}\ct
\aq\partial_1A_1-(g^{00})^2\partial_0A_0\cq 
\nn\\&&\hs
+\frac2{r^2}\at1-\frac{2M}{r}\ct A_1
+\frac{2g^{23}}{r}(\partial_2A_3+\partial_3A_2)
+\partial_1\nabla_jA^j\:,\label{LA1}\\
\Ls A_2&=&
-\frac{2\bar z(1+z\bar z)}{r^2}\partial_3A_2
+\frac{2g^{11}}{r}(\partial_1A_2-\partial_2A_1)
+\partial_2\nabla_jA^j\:,\label{LA2}\\
\Ls A_3&=&
-\frac{2z(1+z\bar z)}{r^2}\partial_3A_3
+\frac{2g^{11}}{r}(\partial_1A_3-\partial_3A_1)
+\partial_3\nabla_jA^j\label{LA3}\:.
\eeq

In order to select the physical degrees of freedom, now we fix the 
gauge $A_0=0$. In this way Eq.~(\ref{LA0}) gives the constraint
\beq 
\nabla_jA^j-2\at\frac{M}{r^2}-\frac{Q^2}{r^3}\ct A_1=
\frac{g^{rr}}{r^2}\partial_r(r^2A_r)
+g^{z\bar z}(\partial_zA_{\bar z}
+\partial_{\bar z}A_z)=0\:,\label{gauge}
\eeq
while Eqs.~(\ref{LA1})-(\ref{LA3}) simplify to
\beq
\Ls A_r&=&-\frac2{r^2}\partial_r(rg^{rr}A_r)\:,\label{LLA1}
\eeq
\beq
\aq \Ls +\frac{2\bar z(1+z\bar z)}{r^2}\partial_{\bar z}\cq A_z
&=&\frac{2g^{rr}}{r}\partial_rA_z
-\frac{2}{r}\at1-\frac{3M}{r}+\frac{2Q^2}{r^2}\ct
\partial_zA_r\:,\label{LLA2}\\
\aq \Ls +\frac{2z(1+z\bar z)}{r^2}\partial_z\cq A_{\bar z}
&=&\frac{2g^{rr}}{r}\partial_rA_{\bar z}
-\frac{2}{r}\at1-\frac{3M}{r}+\frac{2Q^2}{r^2}\ct
\partial_{\bar z}A_r\:.\label{LLA3}
\eeq

It has to be noted that for any function $\psi(t,r,z,\bar z)$ one has
\beq 
\aq \Ls +\frac{2\bar z(1+z\bar z)}{r^2}\partial_{\bar z}\cq 
\partial_z\psi&=&\partial_z \Ls \psi\:,\nn\\
\aq \Ls +\frac{2\bar z(1+z\bar z)}{r^2}\partial_z\cq 
\partial_{\bar z}\psi&=&\partial_{\bar z} \Ls \psi\:.\nn
\eeq
This means that the variables can be easily separated by putting
$A_z=\partial_z\psi$, $A_{\bar z}=\pm\partial_{\bar z}\psi$
(note that in principle one could choose two different functions $\psi$, 
but this is unnecessary, since only the sum enters the other equations).
Now, one can directly verify that two classes of independent 
eigenfunctions $A\equiv(A_t,A_r,A_z,A_{\bar z})$
of Eqs.~(\ref{LLA1})-(\ref{LLA3}), 
satisfying the constraint, Eq.~(\ref{gauge}), can be put in the form
\beq 
A^{(1)}\equiv\at 0,0,\frac1{\sqrt{2l(l+1)\om}}\:\partial_z\Phi\:,
\frac{-1}{\sqrt{2l(l+1)\om}}\:\partial_{\bar z}\Phi\ct\:,
\eeq
\beq 
A^{(2)}\equiv\at 0,\sqrt{\frac{l(l+1)}{2\om^3}}\frac{\Phi}{r^2}\:,
\frac{g^{rr}}{\sqrt{2l(l+1)\om^3}}\:\partial_z\partial_r\Phi\:,
\frac{g^{rr}}{\sqrt{2l(l+1)\om^3}}
\partial_{\bar z}\:\partial_r\Phi\ct\:,
\eeq
where $\Phi(t,r,z,\bar z)=e^{-i\om t}f(r)Y_l^m(z,\bar z)$ 
is a scalar field satisfying the equation
\beq 
\Box\Phi=\frac{2g^{rr}}{r}\:\partial_r\Phi\:.\label{LAFi}
\label{ECS}
\eeq

In the tortoise coordinate 
\beq 
r^*&=&r+\frac{r_+^2}{r_+-r_-}\:\ln(r-r_+)
-\frac{r_-^2}{r_+-r_-}\:\ln(r-r_-)\:,\nn\\
r_\pm&=&M\pm\sqrt{M^2-Q^2}\:,
\nn\eeq
the radial part of the field satisfies the ordinary differential equation 
\beq 
\aq \frac{d^2}{dr^{*2}}+\om^2-V_l(r^*)\cq\:f(r^*)=0\:,
\hs V_l(r^*)=\frac{l(l+1)}{r^2}\:g^{rr}\:,
\eeq
with the normalization property
\beq 
\int |f(r^*)|^2 dr^* =1.
\eeq
$r_+=r_h$ is the radius of the horizon.
The above solutions form a set of orthonormal eigefunctions with respect to
the scalar product \cite{higu92-46-3450,more96-375-54} 
\beq 
\at A^{(1)},A^{(2)}\ct=i\int g^{ij}\at A_i^{*(1)}\partial_t A_j^{(2)}
-\partial_t A_i^{*(1)} A_j^{(2)}\ct
\:g_{z\bar z}g_{rr}\:dz\:d{\bar z}\:dr\:.
\eeq
Note that the more general expression for the scalar product has been 
specialized to our particular case.

\s{One-loop contribution to the entropy}

The (leading) one-loop contribution to the entropy of the black hole due to 
the electromagnetic field can be easily computed using the brick-wall 
method \cite{thoo85-256-727,thoo96-11-4623}. 
The computation is parallel to the 
original one given by 't~Hooft.
In the WKB approximation, the energy spectrum is given by
\beq 
\oint k_l(r^*)\:dr^*=\oint k_l(r)g_{rr}\:dr=
2\pi\hbar n\:,
\hs n\in\N\:,
\eeq
where 
\beq 
k_l(r^*)^2=\om^2-V_l(r^*)
=E^2-\frac{l(l+1)\:g^{rr}}{r^2}=k_l(r)^2\:.
\eeq

Following 't~Hooft, we consider the field in the region $r_h+\ep<r<R$ and
suppose it to satisfy Dirichlet boundary conditions. Then, the number
of eigenstates with energy smaller than $E$ read
\beq 
\nu(E)&=&\frac1\pi\sum_l(2l+1) n=\sum_l(2l+1)
\int_{r_h+\ep}^R\sqrt{E^2-V_l(r)}\:g_{rr}\:dr\nn\\
&\sim&
\frac1\pi\int_0^{\frac{r^2E^2}{g^{rr}}+1/4}\:d\la
\int_{r_h+\ep}^R\sqrt{E^2-\frac{g^{rr}}{r^2}
\at\la-\fr14\ct}\:g_{rr}\:dr\:,
\eeq
where we have put $\la=\hbar^2(l+1/2)^2$ 
and $\hbar^2\sum_l(2l+1)\to\int d\la$.
The extreme of integration in the variable $\la$ is due to the fact that
$k_l(r)$ has to be positive. 
The integration in $\la$ can be performed and so
\beq 
\nu(E)&\sim&\frac2{3\pi}
\int_{r_h+\ep}^R\at E^2+\frac{g^{rr}}{4r^2}\ct^{3/2}
r^2g_{rr}^2\:dr\nn\\
&=&\frac2{3\pi}
\int_{\ep}^{r_h}\frac{r_h^4}{x^2}\at1+\frac{x}{r_h}\ct^4
\at E^2+\frac{x}{4r_h^3(1+x/r_h)^3}\ct^{3/2}\:dx
\nn\\&&\hs\hs+\frac2{3\pi}
\int_{r_h}^{R-r_h}x^2\at1+\frac{r_h}{x}\ct^4
\at E^2+\frac1{4x^2(1+r_h/x)^3}\ct^{3/2}\:dx\nn\\
&\sim& 
\frac{2r_h^6E^3}{3\pi\ep(r_h-r_-)^2}
+\frac{VE^3}{6\pi}+...
\label{dens}
\eeq
where in the latter expression only the leading divergences have been
written down ($V$ is the volume of the spherical box).
The derivative of $\nu(E)$ in Eq.~(\ref{dens}) represents the density 
of states with energy $E$. 

Now for the partition function one easily gets
\beq 
\ln Z&=&-\sum_\nu\ln\at1-e^{-\be E_\nu}\ct
=\be\int_{0}^{\ii}\frac{\nu(E)}{e^{\be E}-1}\:dE\nn\\
&\sim&
\frac{\pi^2V}{90\be^3}+\frac{2\pi^3r_h^6}{45\ep\be^3(r_h-r_-)^2}+...
\eeq 
which agrees with a similar expression in Ref.~\cite{ghos95-357-295}
and reduces to  the expression given in 
Refs.~\cite{thoo85-256-727,thoo96-11-4623}  
in the the limit $Q\to0$, that is $r_h=2M$, $r_-=0$.
The first term on the right hand side of the latter equation  
is the usual one proportional to the volume, while the second
is a divergent contribution due to the presence of the horizon and
is interpreted as a quantum contribution to the black hole entropy
due to the matter field.
Taking into account that we have two independent scalar fields 
both satisfying Eq.~(\ref{ECS}), we finally get
\beq 
S_{RN}=-\at\be\partial_\be-1\ct\ln Z
=\frac{16\pi^3r_h^6}{45\ep\be^3(r_h-r_-)^2}\:.
\eeq
As already anticipated in the introduction, the leading term in the 
one-loop contribution to the entropy due to the electromagnetic
field is exactly twice the one due to the scalar field. 
In our derivation we do not obtain anomalous terms of the kind obtained
for the Rindler case in 
Refs.~\cite{kaba95-453-281,more96-54-7459,more97-55-3552}. 
In any case, as suggested in 
Ref.~\cite{more97-55-3552},   
such terms are non physical and have to be discharged.

All results of this section have a good limit for $Q\to0$ and so they are
valid also for the Schwarzschild black hole, with the simple substitution
$r_h=2M$, $r_-=0$. 

At the equilibrium temperature  $T_H=\frac{r_h-r_-}{4\pi r_h^2}$
the entropy reads
\beq 
S_{T=T_H}=\frac{\sqrt{M^2-Q^2}}{90\ep}
\sim\frac{1}{45}\at\frac{r_h}l\ct^2\:,
\hs l^2=\frac{4r_h^2\ep}{r_h-r_-}\:,
\eeq
where the cutoff parameter $\ep$ has been expressed in terms of the 
proper distance $l$.

\s{Conclusion}
We have written the Maxwell equations in the Reissner-Nordst\"om 
background in terms of a couple of scalar
fields satisfying a Klein-Gordon like equation. 
In this way we have shown that the first quantum correction
to the black hole entropy due to the electromagnetic field, 
which we have computed by using the semi-classical approximation, 
is exactly twice the one which one has for a scalar field. 
This means that the anomalous contributions, which one has  
for the Rindler case \cite{kaba95-453-281,more96-54-7459,more97-55-3552}, 
 are not present in this context.

\ack{It is a pleasure to thank V.~Moretti, L.~Vanzo and S.~Zerbini for
many discussions and suggestions.}

\end{document}